# Water Production Rates from SOHO/SWAN Observations of Comets C/2020 S3 (Erasmus), C/2021 A1 (Leonard) and C/2021 O3 (PanSTARRS)

Short Title: SOHO/SWAN Observations of comets C/2020 S3 (Erasmus), C/2021 A1 (Leonard) and C/2021 O3 (PanSTARRS)

M.R. Combi[1], T. Mäkinen[2], J.-L. Bertaux[3], E. Quémerais[3], and S. Ferron[4]

[1]Dept. of Climate and Space Sciences and Engineering
University of Michigan
2455 Hayward Street
Ann Arbor, MI 48109-2143
United States of America
*Corresponding author: mcombi@umich.edu

[2]Finnish Meteorological Institute, Box 503
SF-00101 Helsinki, FINLAND

[3]LATMOS/IPSL
Université de Versailles Saint-Quentin
11, Boulevard d'Alembert, 78280, Guyancourt, FRANCE

[4]ACRI-st, Sophia-Antipolis, FRANCE

**ABSTRACT**

In 2021 and 2022 the hydrogen comae of three long period comets, C/2020 S3 (Erasmus), C/2021 A1 (Leonard) and C/2021 O3 (PanSTARRS) were observed with the Solar Wind ANisotropies (SWAN) all-sky hydrogen Lyman-alpha camera on the SOlar and Heliosphere Observer (SOHO) satellite. SWAN obtains nearly daily full-sky images of the hydrogen Lyman-$\alpha$ distribution of the interstellar hydrogen as it passes through the solar system yielding information about the solar wind and solar ultraviolet fluxes that eats away at it by ionization and charge exchange. The hydrogen comae of comets, when of sufficient brightness, are also observed. Water production rates have been calculated over time for each of these comets. Of particular interest are comet C/2021 O3 (PanSTARRS) which apparently disintegrated a few days before its perihelion at 0.28 au and C/2021 A1 (Leonard) which also disintegrated beginning about 20 days after its perihelion peak. The behavior of comet C/2020 S3 (Erasmus) was more typical without dramatic fading, but still was asymmetric about perihelion, with a more rapid turn on before perihelion and more extended activity well after perihelion.

**1. INTRODUCTION**

Three long period comets were observed by the Solar Wind Anisotropies (SWAN) all-sky hydrogen Lyman-$\alpha$ camera on the Solar and Heliosphere Observer (SOHO) satellite during 2021 and 2022: C/2020 S3 (Erasmus), C/2021 A1 (Leonard) and C/2021 O3 (PanSTARRS). SWAN's main investigation is to make all-sky images at H Ly$\alpha$ to measure the spatial structure of the depletion of the interstellar hydrogen streaming through the solar system. This provides a global picture of the solar wind as it eats away the interstellar hydrogen (Bertaux et al. 1995). SOHO and SWAN have been operating since launch in December 1995 in a halo orbit around the L1 Sun-Earth Lagrange point. From this location SOHO's *in situ* instruments detect the solar wind and its composition upstream from the Earth and most of its remote sensing instruments image the Sun and surrounding environs in a range of wavelengths. Because water is the main volatile ice in most comets, they are surrounded by a huge atmosphere, or coma, of atomic hydrogen produced by the photodissociation of the water and subsequently by intermediate products OH and $H_2$ (Combi et al. 2004) as water is sublimated from the nucleus and streams outward from the near gravity free nucleus. Moderate to very active comets that are close enough to the Sun produce a hydrogen coma that is bright enough for SWAN to detect and image.

Since January 1996, beginning with the bright comet C/1996 B2 (Hyakutake), well over 70 comets have been observed with SWAN. Including several short period comets that have been observed several times each over 26 years, over 90 individual comet apparitions have been observed and analyzed (Combi et al 2019; Combi 2017, 2020, 2022). Water production rates are calculated using the method described by Mäkinen and Combi (2005). The method accounts for the photodissociation rates, exothermal velocities and partial thermalization of H atoms through collisions with heavy molecules. Table 1 describes the basic orbital parameters and a summary of SWAN observations of the three comets described in this paper. In the tables and figures contained herein, uncertainties, dQ, are 1-sigma uncertainties resulting from the fit of the shape of the model spatial distribution compared with the observation and the underlying interplanetary hydrogen background and noise in the data. Actual uncertainties accounting also for the calibration of the SWAN instrument, that of the solar Lyman-alpha flux obtained from LASP (http://lasp.colorado.edu/lisird/lya/), the model itself and internal parameters, combined with faint field stars not able to be explicitly accounted for are on the order of 30%.

## 2. C/2020 S3 (Erasmus)

Comet C/2020 S3 (Erasmus) was discovered by Nicolas Erasmus of the Asteroid Terrestrial-Impact Last Alert System (ATLAS) on 17 September 2020. The comet reached a perihelion distance of 0.399 au on 12.66 December 2020. According to the IAU Minor Planet Center the original reciprocal semimajor axis was 0.004996, or a semimajor axis of 200 au, indicating that it is an old long period comet in the A'Hearn et al. (1995) classification and definitely not dynamically new. It was likely last through the inner solar system ~2600 years in the past and possibly several times before that. SWAN observed S3 beginning on 28 October 2020 and on most days until 5 December, a week before perihelion. Because of observational constraints it was not observable again until 19 January 2021. It was subsequently observed on most days until 13 February.

Water production rates were calculated from each image with the time-resolved model (TRM), which is described in detail in the paper by Mäkinen and Combi (2005). The observational circumstances and water production rates are given in Table 2. The water production rate is plotted as a function of time in Figure 1 and as a function of heliocentric distance in Figure 2. The water production rate was relatively constant at values between 1 and 2 x $10^{28}$ s$^{-1}$ from 29 October until 12 November, at which time it rose surprisingly rapidly before perihelion, varying as the heliocentric distance, r in au, with an exponent of $r^{-5}$ to a maximum value on 5 December 2020, a few days before perihelion, of 3.3 x $10^{29}$ $s^{-1}$. Once it was observable again after perihelion the value decreased from ~$10^{29}$ $s^{-1}$ to ~5 x $10^{29}$ $s^{-1}$ and was ~5 times larger than at comparable heliocentric distances before perihelion. Since the comet has been through the inner solar system in the past the behavior is likely a seasonal effect rather some blow-off or steadying that is often seen in truly dynamically new comets (Combi et al. 2019; Combi et al. 2013). In a survey of 61 comets observed with SWAN through 2017 Combi et al. (2019) found that for long-period Oort Cloud comets pre-perihelion slopes, or the exponent of the production rate variation with heliocentric distance, generally had few large negative values for comets that are more dynamically new. They were generally limited to values between -3 and -1. For comets that were older, indicating that they had passed through the inner solar system since being initially perturbed from their origin in the Oort cloud, they had a wider variation of slopes with many in the range of -3 to -9 indicating their initial outer layers had been lost and the surfaces more evolved. The ranges of post-perihelion slopes was less distinguishing. For Jupiter family comets some slopes were as steep as -10 to -14 indicating even more evolved surfaces.

Water production rates in Table 2 are consistent with contemporaneous values determined from ground based OH observations by Jehin et al. (2022a, 2022b, 2022c) on November 11, which is shown as the X in Figure 1 and converted to a water production rate using the photochemical yield of 0.84 OH radicals per $H_2O$ molecule (Combi et al. 2004). The use of the empirically determined Haser scale lengths by Jehin et al. for OH with an outflow velocity of 1 km $s^{-1}$ is reasonably consistent with our necessarily more complicated TRM analysis of H and should not introduce inconsistencies much larger than 20% for these observations. Their observation on October 14 is well before the SWAN observations and off the lower left corner of the plot in Figure 1.

**3. Comet C/2021 A1 (Leonard)**

Comet C/2021 A1 (Leonard), hereafter A1, was discovered in images taken at the Mount Lemmon Observatory on 3 January 2021 by G.J. Leonard almost exactly a year before its perihelion of 0.615 au on 3 January 2022. According to the IAU Minor Planets Center comet A1 had an original reciprocal semi-major axis of 0.00052 $au^{-1}$ or a semi-major axis of 1923 au indicating that it is Young Long Period comet using the A'Hearn et al. (1995) classification. This means that it is not dynamically new and has been through the inner solar system in the past about 84000 years ago. Indications are that it, or more accurately its remaining fragment cloud, is now on a slightly hyperbolic orbit never to return according to the Minor Planet Center final orbital parameters.

SWAN obtained images of the hydrogen Lyman-α coma of comet A1 beginning on 1 October 2021. Starting on 2 November SWAN obtained nearly daily images of the comet until 29 December 2021 5 days before perihelion after which it was too close to the Sun in the sky until 11 January 2022 when it then observed almost daily until 3 February 2022. The water production rates as well as the observational circumstances are given in Table 3 and are plotted as a function of time in Figure 3. Figure 3 also shows three water production rates derived from OH measurements by Jehin et al. (2020a, 2020b), again using the photochemical yield of 0.84 OH radicals per $H_2O$ molecule (Combi et al. 2004), which are quite consistent with the SWAN results. The pre-perihelion activity of A1 is remarkably qualitatively similar to S3. The water production rates are relatively constant increasing slowly from the early observations from $1 \times 10^{28}$ $s^{-1}$ until about 20 days to about $2 \times 10^{28}$ $s^{-1}$ before perihelion. At that time, the water production rate increased rapidly until just before perihelion when it reached a value of $\sim 2 \times 10^{29}$ $s^{-1}$ when the comet was too close to the Sun for SWAN to observe. Beginning 10 days after perihelion the production rate was initially similar to the value before perihelion at comparable heliocentric distances, and then dropprd rapidly, like the pre-perihelion part of the orbit, until 20 days after perihelion with a heliocentric distance exponent of -4 to -5, again like S3. However then the water production rate continued to drop dramatically, consistent with the comet having disintegrated rather than leveling off like it had before perihelion.

While disintegration is more common for dynamically new comets, the behavior of A1 possibly means that on previous apparitions the comet did not reach such a small perihelion distance, more likely coming only into the region of the giant planets in the past, where its original dynamically new and distant aphelion orbit could have been perturbed to its current orbit. Unfortunately the refractory/water ice ratio in comets is still not a well-determined value, even after the Rosetta mission results which provided reasonable estimates of the total mass, shape and therefore volume of the nucleus (Pázold et al. 2016) as well as the total water mass lost throughout the mission. Estimates of the ice to dust ratio range from ~1 from results of the total water mass loss (Combi et al. 2020, Biver et al. 2019) to a value of 6±9 in a study of the possible non-escaping and fallback dust ( Fulle et al. 2019).

The total mass of water lost after the beginning of the apparent outburst indicating the breakup of the nucleus at 20 days after perihelion was $2.0 \times 10^9$ kg. We can assume that the total nucleus mass is approximately 2 to 10 times the lost water mass. If we use the nucleus density of 533 kg m$^{-3}$ from the Rosetta result of comet 67P/Churyumov-Gerasimenko (Päzold et al. 2016), then we find an approximate radius of the nucleus between 121 and 192 m just before its disintegration. We can similarly calculate a nucleus size before the entire apparition, assuming that the production continued to vary as approximate $r^{-4}$ throughout the 15-day gap around perihelion. This results in a total water mass loss of $2.1 \times 10^{10}$ kg or an approximate original nucleus radius of between 265 and 461 m.

**5. Comet C/2021 O3 (PanSTARRS)**

Comet C/2021 O3 (PanSTARRS) was discovered on 26 July 2021 by the PanSTARRS NEO Sky Survey (Weryk, 2021) with the 1.8-m Ritchey-Chretien reflector at Haleakala, Hawaii. The comet reached a perihelion distance of 0.287 au on 31 April 2022. The comet's original reciprocal semi-major axis was zero within uncertainties according to the IAU Minor Planet Center and so was a truly dynamically new comet. Because of such a small perihelion distance, hopes were high of a spectacular visual display. SWAN detected the comet on most days from 6 April until 24 April 2022.

Again we used the TRM to calculate water production rates from each of the images, however the fits of the model to nearly all of the images after April 12 were poor, yielding uncertainties that were as large or even larger than the calculated production rate. The water production rates as a function of time from perihelion are shown in Figure 4 and given in Table 4. The water production rate increased from April 6 to a maximum value of $(9.26 \pm 2.49) \times 10^{28}$ s$^{-1}$ on April 10 and then rapidly decreased. Given the large uncertainties and poor model fits after this date it is likely that the comet disintegrated sometime between April 5 and April 10. After that the residual hydrogen coma was detectable but gradually dissipated resulting the poor model fits to what would be expected for the spatial distribution of a typical hydrogen coma continuously being produced from water sublimating from a well-defined nucleus and subsequent outflow and photodissociation. The disintegration was also reported by Zhang et al. (2022) who reported trying to observe the comet on 29 April 2022.

If we take the total mass of water lost during the apparition of $2.0 \times 10^9$ kg, assume the total nucleus mass is between 2 and 4 times the water mass or a refractory/water ratio between 1 and 3, and use the density of 533 kg m$^{-3}$ from the Rosetta result of comet 67P/Churyumov-Gerasimenko (Päzold et al. 2016) and the previous range of estimates of the dust to ice ratio (Combi et al. 2020, Biver et al. 2019, Fulle et al. 2019), we find an approximate radius of between 121 and 192 m just like comet A1. It is therefore quite interesting that both A1 and O3 broke up and disintegrated once they reached a nucleus radius of 121 to 192 m.

## 8. Summary

We describe herein the results of the analysis of the observations of the hydrogen Lyman-alpha comae of three long-period comets observed during the 2020-2022 time period by the SWAN all-sky camera on the SOHO spacecraft. They cover a range from old long-period to young long-period to truly dynamically new comets, using the dynamical classification of A'Hearn et al. (1995).

Comet C/2020 S3 (Erasmus) is an old long period comet with a semi-major axis of 200 au that reached a perihelion distance of 0.399 au on 12.66 December 2020. On the inbound orbit the water production rate was rather flat to within the uncertainties at level of ~1 x 10$^{28}$ s$^{-1}$ from 1.1 au to about 0.88 au and then began to increase with a very steep rise (an exponent of -5) until it reached a production rate of 3.5 x 10$^{29}$ s$^{-1}$ just before perihelion. It could not be observed then until it reached 1 au on the outbound leg at which time the production rate decreased very slowly again out to 1.45 au, but at a level of ~7 times that of the flat level before perihelion.

Comet C/2021 A1 (Leonard) is a young long period comets with a semi-major axis of 1923 au indicating that it left the planetary region of the solar system on its previous apparition about 84000 years ago. Despite not being dynamically new, between 21 and 26 days after its perihelion distance of 0.615 au it had a 40% outburst of activity and then disintegrated. The hydrogen coma was not detectable after another 10 days. The water production rates (Mäkinen et al. 2001) of the well-observed comet C/1999 S4 (Linear) also took 10 days for the production rate to drop this much.

Comet C/2021 O3 (PanSTARRS) is (or was) a truly dynamically new comet with an original reciprocal semi-major axis indistinguishable from barely hyperbolic. O3, or what was left of it, reached a perihelion distance of 0.287 au. It began to disintegrate between 5 and 9 April 2022, and the water production rate dropped by nearly a factor of 10 over the next two weeks.

From the water production rate during the time after the disintegration even for both comets A1 and O3 we estimated the radius of both nuclei was between 121 to 192 m corresponding to refractory to water ice ratios of 1 to 9. This radius is even smaller than the average size found by Jewitt (2022) for a range of long-period comets that have disintegrated or fragmented.

**Acknowledgements:** SOHO is an international mission between ESA and NASA. M. Combi acknowledges support from NASA grant 80NSSC18K1005 from the Solar System Observations Program. T.T. Mäkinen was supported by the Finnish Meteorological Institute (FMI). J.-L. Bertaux and E. Quémerais acknowledge support from CNRS and CNES. We obtained orbital elements from the JPL Horizons web site (http://ssd.jpl.nasa.gov/horizons.cgi). For classification of the dynamical ages of the comets in this paper we used the Minor Planets Center web site https://www.minorplanetcenter.net/db search tool. The composite solar Ly$\alpha$ data were taken from the LASP web site at the University of Colorado (http://lasp.colorado.edu/lisird/lya/). We acknowledge the personnel who have been keeping SOHO and SWAN operational for over 25 years, in particular Dr. Walter Schmidt at FMI. We also thank the two reviewers for their careful reading and helpful suggestions that have improved the paper.

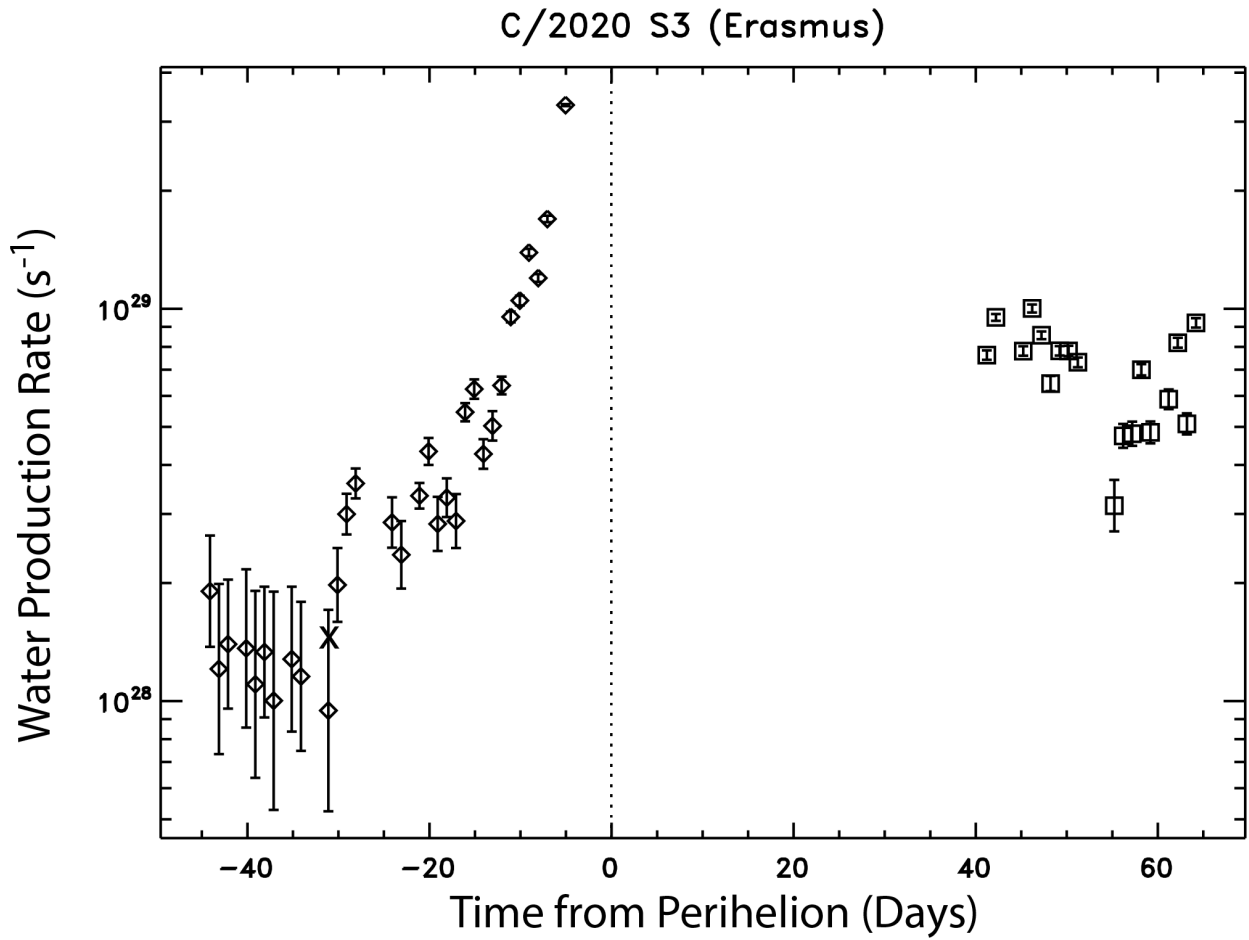

Figure 1. Water production rate of comet C/2020 S3 (Erasmus) as a function of time from perihelion. The points give the water production rate in s$^{-1}$ from single images. The error bars give the 1-σ formal random fitting errors for each value. There is a ~30% uncertainty from the model parameters and calibration. The X shows the water production rate from the OH ground-based measurements of Jehin et al. (2020a, 2020b) at T = -31 days from perihelion.

# C/2020 S3 (Erasmus)

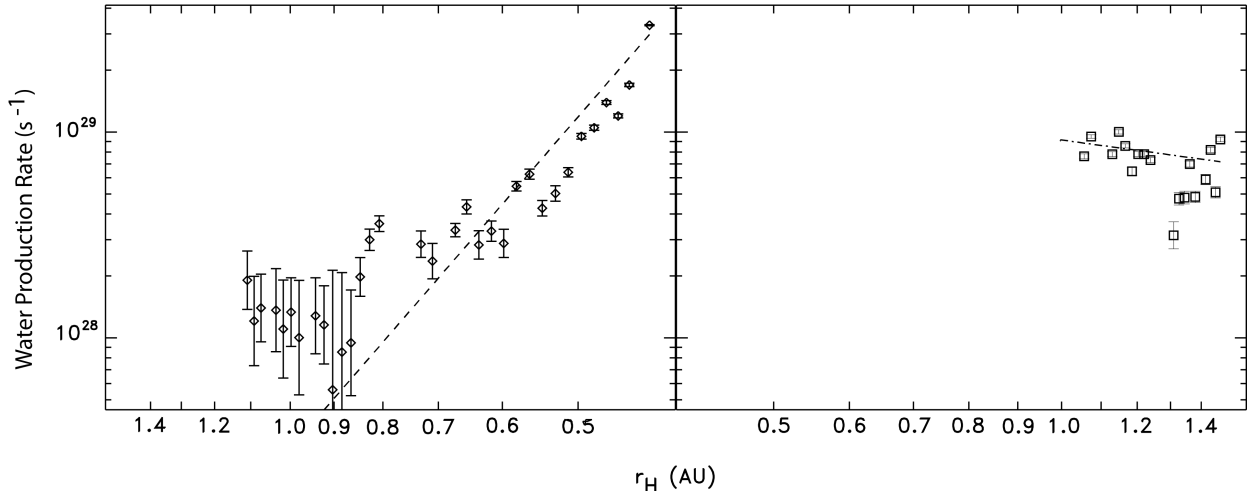

Figure 2. Water production rate in C/2020 S3 (Erasmus) as a function of heliocentric distance ($r_H$). The points give the water production rate in s$^{-1}$ from single images. The error bars give the 1-$\sigma$ formal random fitting errors for each value. There is a ~30% uncertainty from the model parameters and calibration. The dashed line in the left panel gives the variation exponent of -5.4 before perihelion, and the dot-dash line in the right panel gives the variation exponent of -0.6 after perihelion.

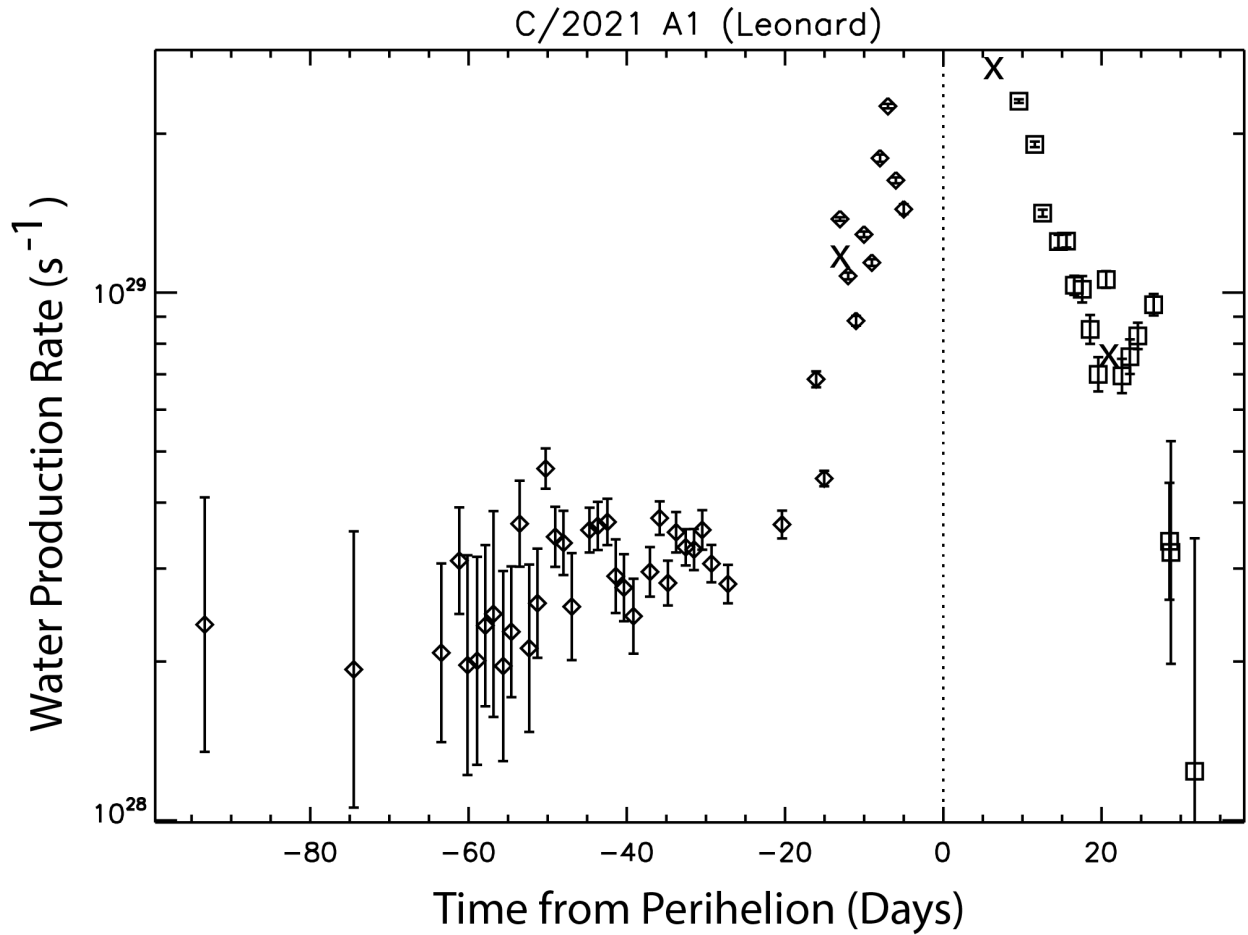

Figure 3. Water production rate in comet C/2021 A1 (Leonard) as a function of time from perihelion. The points give the water production rate in s$^{-1}$ from single images. The error bars give the 1-$\sigma$ formal random fitting errors for each value. There is a ~30% uncertainty from the model parameters and calibration.

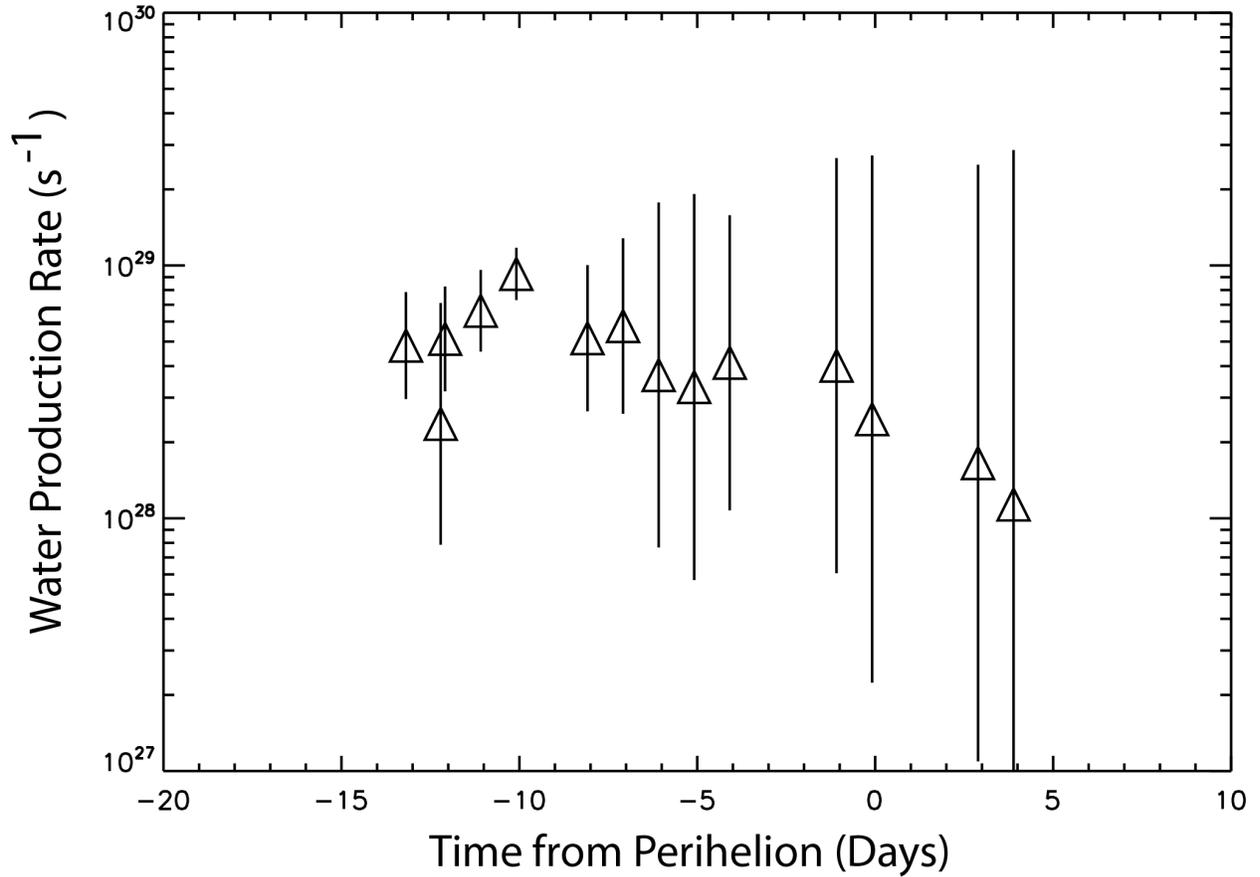

Figure 4. Water production rate in comet C/2021 O3 (PanSTARRS) as a function of time from perihelion and UT date 2020. The points give the water production rate in s$^{-1}$ from single images. The error bars give the 1-σ formal random fitting errors for each value. There is a ~30% uncertainty from the model parameters and calibration.

Table 1

Summary of SOHO/SWAN Observations

| Comet | Perihelion Date | q(au) | Images | $r_H$(au) |
|---|---|---|---|---|
| C/2020 S3 (Erasmus) | 2020 Dec 12.68 | 0.397 | 49 | 0.421 - 1.486 |
| C/2021 A1 (Leonard) | 2022 Jan 03.30 | 0.615 | 61 | 0.625 - 1.843 |
| C/2021 O3 (PanSTARRS) | 2022 Apr 21.05 | 0.287 | 14 | 0.313 - 0.491 |

Notes to Table 1

$r_H$ = heliocentric distance (au) range

q = perihelion distance (au)

Table 2

SOHO/SWAN Observations of C/2020 S3 (Erasmus) and Water Production Rates

| ΔT (Days) | R (au) | Δ (au) | g (s$^{-1}$) | Q (10$^{28}$ s$^{-1}$) | δQ (10$^{28}$ s$^{-1}$) |
|---|---|---|---|---|---|
| −44.107 | 1.109 | 1.187 | 0.001862 | 1.91 | 0.74 |
| −43.123 | 1.091 | 1.172 | 0.001869 | 1.21 | 0.78 |
| −42.123 | 1.073 | 1.157 | 0.001896 | 1.40 | 0.64 |
| −40.123 | 1.035 | 1.130 | 0.001921 | 1.36 | 0.81 |
| −39.123 | 1.017 | 1.117 | 0.001948 | 1.10 | 0.81 |
| −38.123 | 0.998 | 1.105 | 0.001941 | 1.33 | 0.62 |
| −37.106 | 0.979 | 1.093 | 0.001934 | 1.00 | 0.90 |
| −35.123 | 0.941 | 1.073 | 0.001946 | 1.28 | 0.68 |
| −34.106 | 0.922 | 1.063 | 0.001944 | 1.16 | 0.63 |
| −31.106 | 0.864 | 1.040 | 0.001894 | 0.95 | 0.76 |
| −30.106 | 0.845 | 1.034 | 0.001916 | 1.98 | 0.48 |
| −29.106 | 0.826 | 1.029 | 0.001932 | 3.00 | 0.38 |
| −28.105 | 0.807 | 1.025 | 0.001947 | 3.59 | 0.33 |
| −24.105 | 0.700 | 1.018 | 0.001909 | 2.85 | 0.45 |
| −23.098 | 0.710 | 1.018 | 0.001996 | 2.36 | 0.52 |
| −21.098 | 0.672 | 1.022 | 0.002054 | 3.34 | 0.26 |
| −20.098 | 0.654 | 1.026 | 0.002063 | 4.33 | 0.36 |
| −19.098 | 0.635 | 1.030 | 0.002055 | 2.83 | 0.49 |
| −18.098 | 0.616 | 1.036 | 0.002039 | 3.30 | 0.40 |
| −17.078 | 0.598 | 1.043 | 0.002022 | 2.88 | 0.49 |
| −16.078 | 0.580 | 1.050 | 0.002043 | 5.45 | 0.30 |
| −15.078 | 0.562 | 1.059 | 0.002016 | 6.24 | 0.37 |
| −14.077 | 0.545 | 1.068 | 0.001969 | 4.27 | 0.39 |
| −13.077 | 0.528 | 1.078 | 0.001928 | 5.03 | 0.45 |
| −12.073 | 0.512 | 1.090 | 0.001869 | 6.38 | 0.34 |
| −11.074 | 0.496 | 1.102 | 0.001833 | 9.54 | 0.31 |
| −10.074 | 0.481 | 1.114 | 0.001789 | 10.50 | 0.30 |
| −9.074 | 0.467 | 1.128 | 0.001746 | 13.92 | 0.30 |
| −8.051 | 0.454 | 1.143 | 0.001719 | 11.98 | 0.27 |
| −7.051 | 0.442 | 1.158 | 0.001669 | 16.95 | 0.33 |
| −5.051 | 0.421 | 1.189 | 0.001588 | 33.07 | 0.20 |
| 41.215 | 1.056 | 1.988 | 0.001781 | 7.62 | 0.21 |
| 42.215 | 1.074 | 2.006 | 0.001769 | 9.51 | 0.19 |

| DT | r | Δ | g | Q | δQ |
|---|---|---|---|---|---|
| 45.215 | 1.130 | 2.063 | 0.001697 | 7.81 | 0.22 |
| 46.191 | 1.148 | 2.081 | 0.001698 | 10.02 | 0.24 |
| 47.215 | 1.166 | 2.100 | 0.001691 | 8.57 | 0.20 |
| 48.215 | 1.185 | 2.119 | 0.001679 | 6.45 | 0.29 |
| 49.215 | 1.203 | 2.138 | 0.001674 | 7.82 | 0.22 |
| 50.191 | 1.220 | 2.156 | 0.001671 | 7.81 | 0.24 |
| 51.215 | 1.239 | 2.175 | 0.001715 | 7.31 | 0.21 |
| 55.216 | 1.310 | 2.249 | 0.001681 | 3.15 | 0.51 |
| 56.191 | 1.327 | 2.267 | 0.001699 | 4.75 | 0.35 |
| 57.191 | 1.345 | 2.286 | 0.001692 | 4.80 | 0.36 |
| 58.191 | 1.362 | 2.304 | 0.001699 | 6.99 | 0.24 |
| 59.191 | 1.380 | 2.322 | 0.001712 | 4.84 | 0.32 |
| 61.190 | 1.415 | 2.358 | 0.001712 | 5.88 | 0.35 |
| 62.190 | 1.432 | 2.376 | 0.001721 | 8.20 | 0.24 |
| 63.190 | 1.449 | 2.394 | 0.001708 | 5.09 | 0.33 |
| 64.190 | 1.466 | 2.412 | 0.001731 | 9.21 | 0.26 |

Notes. DT (Days from Perihelion December 12.68, 2020)

r: Heliocentric distance (au)

Δ: Comet-SOHO distance (au)

g: Solar Lyman-α g-factor (photons s$^{-1}$) at 1 au

Q: Water production rates for each image (s$^{-1}$)

δQ: internal 1-sigma uncertainties

Table 3

SOHO/SWAN Observations of C/2020 A1 (Leonard) and Water Production Rates

| $\Delta T$ (Days) | r (au) | $\Delta$ (au) | g (s$^{-1}$) | Q (10$^{28}$ s$^{-1}$) | $\delta Q$ (10$^{28}$ s$^{-1}$) |
|---|---|---|---|---|---|
| −93.332 | 1.843 | 2.406 | 0.001949 | 2.35 | 1.74 |
| −74.503 | 1.563 | 1.893 | 0.001794 | 1.93 | 1.60 |
| −63.432 | 1.393 | 1.551 | 0.002044 | 2.08 | 0.99 |
| −61.163 | 1.358 | 1.478 | 0.001942 | 3.10 | 0.81 |
| −60.121 | 1.342 | 1.443 | 0.001947 | 1.97 | 1.21 |
| −58.913 | 1.324 | 1.404 | 0.001943 | 2.01 | 1.15 |
| −57.871 | 1.308 | 1.369 | 0.001925 | 2.34 | 0.99 |
| −56.848 | 1.292 | 1.335 | 0.001897 | 2.46 | 1.39 |
| −55.621 | 1.273 | 1.293 | 0.001926 | 1.96 | 1.01 |
| −54.580 | 1.257 | 1.258 | 0.001902 | 2.28 | 0.75 |
| −53.538 | 1.240 | 1.222 | 0.001879 | 3.65 | 0.75 |
| −52.330 | 1.222 | 1.181 | 0.001902 | 2.12 | 0.93 |
| −51.288 | 1.206 | 1.145 | 0.001855 | 2.58 | 0.69 |
| −50.247 | 1.189 | 1.109 | 0.001864 | 4.64 | 0.43 |
| −49.038 | 1.171 | 1.067 | 0.001873 | 3.45 | 0.48 |
| −47.997 | 1.155 | 1.030 | 0.001859 | 3.36 | 0.50 |
| −46.955 | 1.138 | 0.993 | 0.001849 | 2.54 | 0.67 |
| −44.705 | 1.104 | 0.913 | 0.001867 | 3.55 | 0.36 |
| −43.664 | 1.088 | 0.876 | 0.001880 | 3.61 | 0.40 |
| −42.445 | 1.069 | 0.832 | 0.001916 | 3.68 | 0.39 |
| −41.403 | 1.053 | 0.795 | 0.001932 | 2.90 | 0.51 |
| −40.361 | 1.037 | 0.757 | 0.001930 | 2.76 | 0.44 |
| −39.153 | 1.019 | 0.714 | 0.001959 | 2.44 | 0.43 |
| −37.070 | 0.987 | 0.639 | 0.001938 | 2.96 | 0.34 |
| −35.843 | 0.969 | 0.595 | 0.001884 | 3.74 | 0.28 |
| −34.801 | 0.953 | 0.557 | 0.001879 | 2.82 | 0.29 |
| −33.760 | 0.938 | 0.520 | 0.001861 | 3.51 | 0.32 |
| −32.551 | 0.920 | 0.478 | 0.001819 | 3.29 | 0.27 |
| −31.509 | 0.905 | 0.442 | 0.001801 | 3.26 | 0.31 |
| −30.467 | 0.890 | 0.407 | 0.001800 | 3.55 | 0.32 |
| −29.301 | 0.873 | 0.369 | 0.001781 | 3.07 | 0.26 |
| −27.217 | 0.844 | 0.307 | 0.001732 | 2.81 | 0.25 |
| −20.377 | 0.755 | 0.234 | 0.001937 | 3.64 | 0.23 |

| DT | r | Δ | g | Q | δQ |
|---|---|---|---|---|---|
| -16.072 | 0.707 | 0.332 | 0.001932 | 6.85 | 0.24 |
| -15.054 | 0.697 | 0.365 | 0.001921 | 4.44 | 0.15 |
| -13.051 | 0.678 | 0.433 | 0.001863 | 13.79 | 0.10 |
| -12.028 | 0.669 | 0.470 | 0.001808 | 10.75 | 0.16 |
| -11.029 | 0.661 | 0.506 | 0.001757 | 8.84 | 0.19 |
| -10.023 | 0.653 | 0.544 | 0.001694 | 12.89 | 0.16 |
| -9.023 | 0.646 | 0.581 | 0.001650 | 11.40 | 0.17 |
| -7.999 | 0.640 | 0.620 | 0.001603 | 17.97 | 0.29 |
| -7.000 | 0.634 | 0.658 | 0.001564 | 22.56 | 0.22 |
| -5.999 | 0.629 | 0.696 | 0.001557 | 16.32 | 0.21 |
| -5.000 | 0.625 | 0.734 | 0.001577 | 14.39 | 0.32 |
| 9.556 | 0.650 | 1.246 | 0.001738 | 23.06 | 0.18 |
| 11.556 | 0.665 | 1.307 | 0.001721 | 19.09 | 0.23 |
| 12.556 | 0.673 | 1.336 | 0.001720 | 14.14 | 0.22 |
| 14.556 | 0.692 | 1.392 | 0.001673 | 12.50 | 0.37 |
| 15.556 | 0.702 | 1.419 | 0.001664 | 12.52 | 0.36 |
| 16.556 | 0.712 | 1.445 | 0.001693 | 10.32 | 0.44 |
| 17.556 | 0.723 | 1.471 | 0.001728 | 10.15 | 0.60 |
| 18.556 | 0.734 | 1.496 | 0.001745 | 8.52 | 0.55 |
| 19.582 | 0.746 | 1.521 | 0.001765 | 7.00 | 0.54 |
| 20.582 | 0.758 | 1.544 | 0.001825 | 10.59 | 0.39 |
| 22.582 | 0.783 | 1.589 | 0.001904 | 6.95 | 0.54 |
| 23.583 | 0.796 | 1.610 | 0.001897 | 7.56 | 0.59 |
| 24.583 | 0.809 | 1.631 | 0.001910 | 8.28 | 0.49 |
| 26.583 | 0.836 | 1.670 | 0.001942 | 9.48 | 0.45 |
| 28.583 | 0.863 | 1.707 | 0.001929 | 3.38 | 0.98 |
| 28.765 | 0.866 | 1.710 | 0.001932 | 3.22 | 2.01 |
| 31.765 | 0.909 | 1.760 | 0.001916 | 1.24 | 2.19 |

Notes. DT (Days from Perihelion January 3.30, 2022)

r: Heliocentric distance (au)

Δ: Comet-SOHO distance (au)

g: Solar Lyman-α g-factor (photons s$^{-1}$) at 1 au

Q: Water production rates for each image (s$^{-1}$)

δQ: internal 1-sigma uncertainties

Table 4

SOHO/SWAN Observations of C/2021 O3 (PanSTARRS) and Water Production Rates

| $\Delta T$ (Days) | r (au) | $\Delta$ (au) | g ($s^{-1}$) | Q ($10^{28}$ $s^{-1}$) | $\delta Q$ ($10^{28}$ $s^{-1}$) |
|---:|---:|---:|---:|---:|---:|
| -13.186 | 0.491 | 1.420 | 0.002312 | 4.82 | 3.02 |
| -12.212 | 0.469 | 1.391 | 0.002311 | 2.36 | 4.74 |
| -12.086 | 0.467 | 1.387 | 0.002311 | 5.13 | 3.14 |
| -11.087 | 0.445 | 1.356 | 0.002287 | 6.62 | 2.98 |
| -10.086 | 0.423 | 1.325 | 0.002262 | 9.26 | 2.49 |
| -8.086 | 0.382 | 1.259 | 0.002185 | 5.16 | 4.88 |
| -7.087 | 0.363 | 1.225 | 0.002132 | 5.76 | 7.04 |
| -6.086 | 0.346 | 1.189 | 0.002077 | 3.69 | 14.05 |
| -5.087 | 0.329 | 1.153 | 0.001998 | 3.31 | 15.87 |
| -4.087 | 0.315 | 1.115 | 0.001905 | 4.12 | 11.68 |
| -1.087 | 0.289 | 0.998 | 0.001688 | 4.01 | 22.56 |
| -0.087 | 0.287 | 0.958 | 0.001657 | 2.47 | 24.74 |
| 2.887 | 0.302 | 0.846 | 0.001723 | 1.65 | 23.36 |
| 3.886 | 0.313 | 0.812 | 0.001785 | 1.13 | 27.48 |

Notes. DT (Days from Perihelion April 21.05, 2022)

r: Heliocentric distance (au)

$\Delta$: Comet-SOHO distance (au)

g: Solar Lyman-$\alpha$ g-factor (photons $s^{-1}$) at 1 au

Q: Water production rates for each image ($s^{-1}$)

$\delta Q$: internal 1-sigma uncertainties